\documentclass[twocolumn]{aastex62}
\usepackage{natbib}
\usepackage{graphicx}
\usepackage[caption=false]{subfig}

\begin{document}

\title{Impacts of Nuclear Reaction Rate Uncertainties on Sub-Chandrasekhar-Mass White Dwarf Detonations}
\author{Thomas Fitzpatrick}
\affil{Department of Astronomy and Theoretical Astrophysics Center, University of California, Berkeley, CA, USA}
\author{Ken J. Shen}
\affil{Department of Astronomy and Theoretical Astrophysics Center, University of California, Berkeley, CA, USA}

\shortauthors{Fitzpatrick et al. }
\correspondingauthor{Thomas Fitzpatrick}
\email{thomas.c.fitzpatrick@berkeley.edu}

\begin{abstract}

The precise nature of Type Ia supernova (SN Ia) progenitors remains a mystery, but the detonation of a sub-Chandrasekhar-mass white dwarf (WD) has become a promising candidate. There is a growing body of work suggesting that the carbon core detonation of a sub-Chandrasekhar-mass WD can be triggered by the detonation of a helium shell accreted from a companion WD, through either inward shock convergence near the center or direct edge-lit detonation. This ``double-detonation" SN Ia can be triggered by a small helium shell and is therefore well approximated by the detonation of a bare carbon-oxygen white dwarf (C/O WD). The impacts of uncertainties in experimentally and theoretically determined nuclear reaction rates on nucleosynthesis in the detonations of sub-Chandrasekhar-mass WDs have not yet been fully explored. We investigate the sensitivity of this model to nuclear reaction rate uncertainties to better constrain the nucleosynthetic yields resulting from these phenomena and identify the nuclear reaction rates whose uncertainties have the most significant impacts on nucleosynthesis. We find that the chemical abundances, and in particular those of the iron-group elements, are relatively insensitive to these nuclear reaction rate uncertainties.
 \end{abstract}

\section{Introduction}

\par Type Ia supernovae are useful cosmological tools as standardizable candles, and this feature was essential in discovering the accelerated expansion of the Universe \citep{1998AJ....116.1009R,1999ApJ...517..565P}. The role of SNe Ia in galactic nucleosynthesis has been thoroughly studied and they are known to contribute significantly to the production of iron-group and intermediate-mass elements crucial to the chemical evolution of the Galaxy \citep{1995ApJS...98..617T}. However, many fundamental questions regarding SNe Ia persist, including the precise nature of the progenitor scenario leading to their explosions.

\par There is a consensus that SNe Ia occur when a WD in a close binary system undergoes explosive burning initiated by mass transfer from either a non-degenerate or degenerate companion star \citep{2014ARA&A..52..107M}. The WD mass distribution produces a typical mass for a single non-rotating white dwarf of $\approx 0.6M_\odot$, which alone, is not vulnerable to nuclear ignition \citep{1998A&A...338..563H}. In the historically favored Chandrasekhar-mass ($M_{\text{Ch}}$) WD scenario, a WD must be accompanied by a companion star from which to accrete mass; the accumulation of this mass leads to increased temperature and density in the core, eventually resulting in nuclear ignition and thermonuclear explosion when the WD nears the Chandrasekhar mass limit. In the single-degenerate case, a C/O WD approaches the Chandrasekhar limit through stable mass accretion from a non-degenerate hydrogen or helium burning companion star \citep{1973ApJ...186.1007W}. In the double-degenerate case, the unstable coalescence of two C/O WDs results in the combined mass of the system exceeding the Chandrasekhar limit \citep{1984ApJ...277..355W}. For all SNe Ia progenitor scenarios, it has been both theoretically and observationally determined that the light curves are dominated by the radioactive decay of newly synthesized $^{56}$Ni in the $^{56}$Ni $\rightarrow$ $^{56}$Co $\rightarrow$ $^{56}$Fe reaction chain \citep{1962PhDT........25P, doi:10.1139/p67-184, 1969ApJ...157..623C}.

\par Alternative to these $M_{\rm Ch}$ scenarios, a class of ``double-detonation" scenarios involving a sub-Chandrasekhar-mass C/O WD has become a promising SN Ia progenitor scenario. The ``double-detonation" scenario involves the carbon-core detonation of a sub-Chandrasekhar-mass (sub-$M_{\text{Ch}}$) C/O WD ignited by the detonation of an accreted helium shell acquired from a companion star, either through inward shock propagation leading to shock convergence near the center of the WD \citep{1990ApJ...354L..53L,1994ApJ...423..371W,2014ApJ...785...61S}, or through direct edge-lit detonation \citep{1982ApJ...257..780N, 1982ApJ...253..798N}. The earliest models of this scenario included a thick helium shell ($\sim 0.1M_\odot$) accreted from a non-degenerate companion star. The detonation of the thick helium shells in these early models yielded an overproduction of iron-group elements including $^{56}$Ni \citep{1996ApJ...457..500H, 1997ApJ...485..812N} and yielded redder spectra than those typically associated to SNe Ia due to Fe-group line blanketing from the ashes of the nuclear burning of these shells\footnote{The recent discovery of SN 2018byg, a thermonuclear transient with observational properties well explained by the detonation of a massive He shell on a sub-$M_{\text{Ch}}$ WD, supports the findings of these earlier models \citep{De_2019}.} \citep{1997ApJ...485..812N, 2010ApJ...719.1067K,2011ApJ...734...38W}.

\par Recent work has resolved this issue, revealing that accretion from degenerate companion WDs yield significantly smaller helium shells, and that the minimum mass helium shell required to trigger a double detonation is much smaller than previously thought \citep{2007ApJ...662L..95B, 2007A&A...476.1133F, 2010A&A...514A..53F, 2014ApJ...797...46S}. A recent study analyzing \emph{Gaia}'s second data release has uncovered hypervelocity WDs that may be the surviving degenerate companions shot out following such a double-degenerate double-detonation SN Ia at extreme velocities equal to their pre-SN orbital velocities \citep{2018ApJ...865...15S}. In addition, recent radiative transfer calculations have shown that the observational signatures from the bulk of observed SNe Ia can be reproduced by the double detonation of sub-$M_{\text{Ch}}$ WDs when non-local thermodynamic equilibrium and multi-dimensional effects are taken into account \citep{2021ApJ...909L..18S, 2021arXiv210812435S}.

\par The impacts of nuclear reaction rate uncertainties have yet to be thoroughly studied in the context of sub-Chandrasekhar-mass progenitor scenarios. The bulk of previous work studying the impacts of nuclear reaction rate uncertainties on the nucleosynthesis of SNe Ia has focused on the W-7 model \citep{1984ApJ...286..644N}, and other near-Chandrasekhar-mass models \citep{2012PhRvC..85e5805B,2013A&A...557A...3P}. Recent studies that have included sub-$M_{\text{Ch}}$ models in their investigations of reaction rate uncertainties have been limited in their scope; focusing on either one isotope \citep{2020MNRAS.499.4725K}, a set of four reactions involving $^{12}$C and $^{16}$O \citep{2019MNRAS.482.4346B}, or on the rates of electron capture reactions \citep{2019A&A...624A.139B}.

\par Motivated by recent observational evidence supporting the sub-$M_{\text{Ch}}$ double detonation model and the need for more detailed nucleosynthetic data, we explore the impacts of nuclear reaction rate uncertainties on the nucleosynthetic yields in the detonations of bare $1.0 \, M_\odot$ C/O WDs.

\section{Numerical Methods}
\par Here we describe our methods for post-processing the one-dimensional hydrodynamic simulation of the detonation of a $1.0 \, M_\odot$ C/O WD described in \citet{2018ApJ...854...52S} used for our study. The explosion of a bare $1.0 \, M_\odot$ C/O WD and the calculation of the nucleosynthetic yield is done in multiple steps. Our choice of WD mass is motivated by the results of \citet{2018ApJ...854...52S}, who find that sufficient $^{56}$Ni is produced in the detonation of a $1.0 \, M_\odot$ C/O WD to power a median brightness Type Ia light curve through the $^{56}$Ni $\rightarrow$ $^{56}$Co $\rightarrow$ $^{56}$Fe decay chain.

\subsection{Detonation Model and Post-Processing}
\par We use the explosion model for a $1.0 \, M_\odot$ C/O WD described in \citet{2018ApJ...854...52S}. The initial conditions for a bare $1.0 \, M_\odot$ C/O WD were calculated using the stellar evolution code \texttt{MESA}, which solves the fully coupled equations of stellar structure and composition simultaneously \citep{2011ApJS..192....3P, 2013ApJS..208....4P, 2015ApJS..220...15P, 2018ApJS..234...34P}. The density profile generated in \texttt{MESA} is used as the initial condition for a \texttt{FLASH} simulation where the detonation is ignited and evolved \citep{2000ApJS..131..273F, 2009arXiv0903.4875D}.  We adopt a WD composition of equal parts $^{12}$C  and $^{16}$O by mass and approximate a solar metallicity ($Z_\odot$) by including $^{22}$Ne and $^{56}$Fe at mass fractions of $X_{22\text{Ne}} = 0.01$ and $X_{56\text{Fe}} = 0.1X_{\text{22Ne}}$, respectively. The 41-isotope nuclear physics network employed in the hydrodynamic simulation, consisting of 190 interlinking reactions from JINA's REACLIB \citep{2010ApJS..189..240C}, is sufficient to accurately calculate energy release but is not adequate for calculations of isotopic abundances \citep{2018arXiv180607820M}. To accurately determine isotopic abundances, tracer particles tracking the radius, velocity, density, and temperature  were included in the explosion model for post-processing, separated uniformly throughout the WD every $5\times10^{6}$cm for a total of 120 zones. We post-process the thermodynamic histories captured by the inner $107$ zones of the explosion model. We ignore nucleosynthetic contributions from the remaining zones as the thermodynamic trajectories are untrustworthy due to the reverse shockwave that travels back through them, but the nucleosynthetic contributions to final isotopic abundances from these zones are negligible. Further details about the numerical setup of the hydrodynamic simulation we use for our study including the detonation procedure and detonation broadening scheme are explained in \citet{2018ApJ...854...52S}.

\par The hydrodynamic simulation is evolved  for $10 \, $s, and we then perform all nucleosynthetic calculations by post-processing the tracer particles' thermodynamic histories using \texttt{MESA}'s one zone burner. To accurately determine nucleosynthetic yields, we couple an extended 205-isotope nuclear reaction network to the density and temperature histories of our tracer particles composed of neutrons, $^{1-2}$H, $^{3-4}$He, $^{6-7}$Li, $^{7,9-10}$Be, $^{8,10-11}$B, $^{12-13}$C, $^{13-16}$N, $^{15-19}$O, $^{17-20}$F, $^{19-23}$Ne, $^{21-24}$Na, $^{23-27}$Mg, $^{25-28}$Al, $^{27-33}$Si, $^{30-34}$P, $^{31-37}$S, $^{35-38}$Cl, $^{35-41}$Ar, $^{39-44}$K, $^{39-49}$Ca, $^{43-51}$Sc, $^{43-54}$Ti, $^{47-56}$V, $^{47-58}$Cr, $^{51-56}$Mn, $^{51-62}$Fe, $^{54-62}$Co, $^{54-62}$Ni, $^{58-66}$Cu, $^{59-66}$Zn, $^{59-66}$Ga, and $^{59-66}$Ge. We include 942 interlinking nuclear reactions, along with their corresponding reverse processes, including $(p,\gamma)$, $(\alpha, \gamma)$ $(n, \gamma)$ $(p, n)$, $(\alpha, n)$, and $(\alpha, p)$ reactions. Nuclear reaction rates are determined from JINA's REACLIB \citep{2010ApJS..189..240C}.

\subsection{Nuclear Reaction Rate Variation}

\par Here we describe our procedure for investigating the extent to which uncertainties in experimentally and theoretically determined nuclear reaction rates impact the nucleosynthetic yields of our sub-$M_{\text{Ch}}$ explosion model. Analogous calculations have been performed by \citet{2012PhRvC..85e5805B} and \citet{2013A&A...557A...3P}, who studied the impacts of nuclear reaction rate uncertainties in the context of the standard $M_{\text{Ch}}$ WD Type Ia supernova models including the delayed detonation model and the carbon deflagration W7 model of \citet{1984ApJ...286..644N} and \citet{1986A&A...158...17T}. We employ a similar approach including the individual normalization of all input nuclear reaction rates by constant factors.

\par We post-process the thermodynamic history of the simulation by coupling the extended nuclear network described above to the density and temperature histories of our tracer particles, individually normalizing the rate of each nuclear reaction in our standard network by two constant factors ($10.0$ and $0.1$ times the standard rate). Note that in order to maintain a detailed balance, for each normalization factor we apply to a reaction rate we apply the same factor to its corresponding reverse process. For each normalization of an individual nuclear reaction rate and its corresponding reverse process, we repeat all nucleosynthetic calculations and determine the final yields of all isotopes in our network $10 \,$s after explosion. To determine the final isotopic abundances we sum the mass-weighted contributions from each of the 107 post-processed zones of our simulation. Performing all nucleosynthetic calculations for all normalizations of the reaction rates, we obtain 1884 sets of post-processed nucleosynthetic results.

\par We compare the nuclear yields produced in the individual normalization of reaction rates by a factor of ten up and down to the nuclear yields generated using standard rates by calculating the relative change between the yields of each isotope in our network for each variation of a reaction rate. The relative change in isotopic abundances is calculated as

\begin{equation}
\text{Relative Change} = \frac{X_{\text{normalized}} - X_{\text{standard}} }{X_{\text{standard}}}
\end{equation}
where $X_{\text{normalized}}$ is the mass fraction of an isotope generated by the normalization of an individual reaction rate, and $X_{\text{standard}}$ is the mass fraction of that same isotope produced using the default rates of our standard nuclear network.

\par In our analysis, we focus our attention on nuclear reaction rates whose uncertainties have the most pronounced effects on nuclear yields. To identify individual reaction rates which have the most significant impact on nuclear abundances, we present only isotopes that achieve an initial abundance of at least $10^{-4}M_\odot$ in the baseline model, and only reaction rates that impact the abundance of an isotope in our network by at least $20\%$ when normalized.

\section{Nucleosynthetic Results} \label{Nucleosynthetic Results}

\par In this section, we show the nucleosynthetic yields for a selection of isotopes in our network and analyze the impact of nuclear reaction rate uncertainties on the yields of these isotopes and the important isotopic ratios $^{57}$Ni/$^{56}$Ni and $^{55}$Co/$^{57}$Ni. The nucleosynthetic yields of all isotopes in our study are calculated $10 \,$s after the start of the explosion, and are not the final stable isotopic yields unless otherwise specified. The value of these ratios can be inferred through observations of late-time SN Ia light curves, and at these late times nucleosynthetic effects of near-$M_{\text{Ch}}$ and sub-$M_{\text{Ch}}$ progenitor scenarios result in differences in the photometric evolution of the light curves. The values of these ratios are therefore useful diagnostic tools in breaking the degeneracy between near-$M_{\text{Ch}}$ and sub-$M_{\text{Ch}}$ progenitor models \citep{2012ApJ...750L..19R, 2014ApJ...796L..26K, 2016ApJ...819...31G, 2017ApJ...841...48S}. We also discuss the impact of reaction rate uncertainties on the value of the Mn to Fe ratio, the value of which can also be used to discriminate between these two scenarios \citep{2013A&A...559L...5S}.

\subsection{Isotopic Yields}

\par To determine the final nucleosynthetic yields we sum over the mass-weighted contributions from each of the 107 post-processed zones of the simulation.  The final isotopic yields of the 32 isotopes selected for our analysis generated using standard nuclear reaction rates are shown in Table \ref{baseline_table}.

\par Again, we restrict our analysis to only isotopes that achieved an abundance of at least $10^{-4}M_\odot$ in our reference model using standard nuclear reaction rates and reactions which impact the abundance of any of these isotopes by at least $20\%$. $^{27}$Al and $^{31}$P do not meet the $10^{-4}M_\odot$ mass criterion for our analysis using standard nuclear reaction rates, but do exceed this criterion in certain cases of a reaction rate variation and we have included both of these isotopes in our analysis: when the rates of  $^{23}$Na($p$, $\gamma$)$^{24}$Mg, $^{20}$Ne($\alpha$, $\gamma$)$^{24}$Mg, and $^{23}$Na($\alpha$, $p$) $^{26}$Mg were normalized by a factor of 10.0, and when the rates of $^{27}$Al($p$, $\gamma$)$^{28}$Si, $^{28}$Si($\alpha$, $\gamma$)$^{32}$S, and $^{28}$Si($\alpha$, $p$)$^{31}$P were normalized by a factor of 0.1, the yields of $^{27}$Al and $^{31}$P were increased past our $10^{-4}M_\odot$ criterion.  Note that the nucleosynthetic yields of the highest mass elements at the end of our nuclear network are uncertain because elements with $Z\geq31$ are neglected.

{\setlength{\tabcolsep}{.25em}
\begin{table}[bht]
\begin{center}
\begin{tabular}{cr|cr}
\hline \hline
Isotope   & \begin{tabular}[c]{@{}r@{}}Ejected Mass \\  ($M_\odot$)\end{tabular} & Isotope   & \begin{tabular}[c]{@{}r@{}}Ejected Mass\\  ($M_\odot$)\end{tabular} \\ \hline \hline
$^{4}$He  & $5.01\times 10^{-3}$                                                 &  $^{40}$Ca & $1.96 \times 10^{-2}$                                                \\
$^{12}$C  & $7.74 \times 10^{-4}$                                                & $^{48}$Cr & $4.14 \times 10^{-4}$                                              \\
$^{16}$O  & $4.92 \times 10^{-2}$                                                & $^{50}$Cr & $1.07 \times 10^{-4}$                                                       \\
$^{20}$Ne & $4.89 \times 10^{-4}$                                                & $^{52}$Fe & $8.89 \times 10^{-3}$                                               \\
$^{24}$Mg & $1.11 \times 10^{-3}$                                                & $^{53}$Fe & $5.52 \times 10^{-4}$                                               \\
$^{27}$Al & $6.46 \times 10^{-5}$                                                & $^{54}$Fe & $1.12 \times 10^{-2}$                                               \\
$^{28}$Si & $1.89 \times 10^{-1}$                                                &  $^{55}$Co & $2.72 \times 10^{-3}$                                                                                                      \\
$^{29}$Si & $1.29 \times 10^{-4}$                                                & $^{56}$Ni & $5.55 \times 10^{-1}$                                            \\
$^{30}$Si & $1.21 \times 10^{-4}$                                                &  $^{57}$Ni & $8.63 \times 10^{-3}$                                              \\
$^{31}$P  & $8.62 \times 10^{-5}$                                                &  $^{58}$Ni & $6.64 \times 10^{-3}$                                              \\
$^{32}$S  & $1.07 \times 10^{-1}$                                                &  $^{59}$Cu & $2.61 \times 10^{-4}$                                               \\
$^{33}$S  & $1.27 \times 10^{-4}$                                                &  $^{60}$Cu & $1.67 \times 10^{-4}$                                            \\
$^{34}$S  & $7.01 \times 10^{-4}$                                                &  $^{60}$Zn & $8.37 \times 10^{-3}$                                              \\
$^{36}$Ar & $2.09 \times 10^{-2}$                                                &  $^{61}$Zn & $2.33 \times 10^{-4}$                                               \\
$^{38}$Ar & $4.66 \times 10^{-4}$                                                &  $^{62}$Zn & $1.23 \times 10^{-3}$                                               \\ \hline \hline

\end{tabular}
\end{center}
\caption{Nucleosynthetic yields generated by the $1.0 \, M_{\odot}$ WD detonation model using standard nuclear reaction rates. We present the results for a selection of 32 important isotopes from our 205-isotope nuclear reaction network that achieve an abundance of at least $10^{-4}M_{\odot}$.}
\label{baseline_table}
\end{table}}

\subsection{Sensitivity to Reaction Rate Normalization}

\begin{figure*}
\begin{center}

\subfloat[]{%
  \includegraphics[scale=.57]{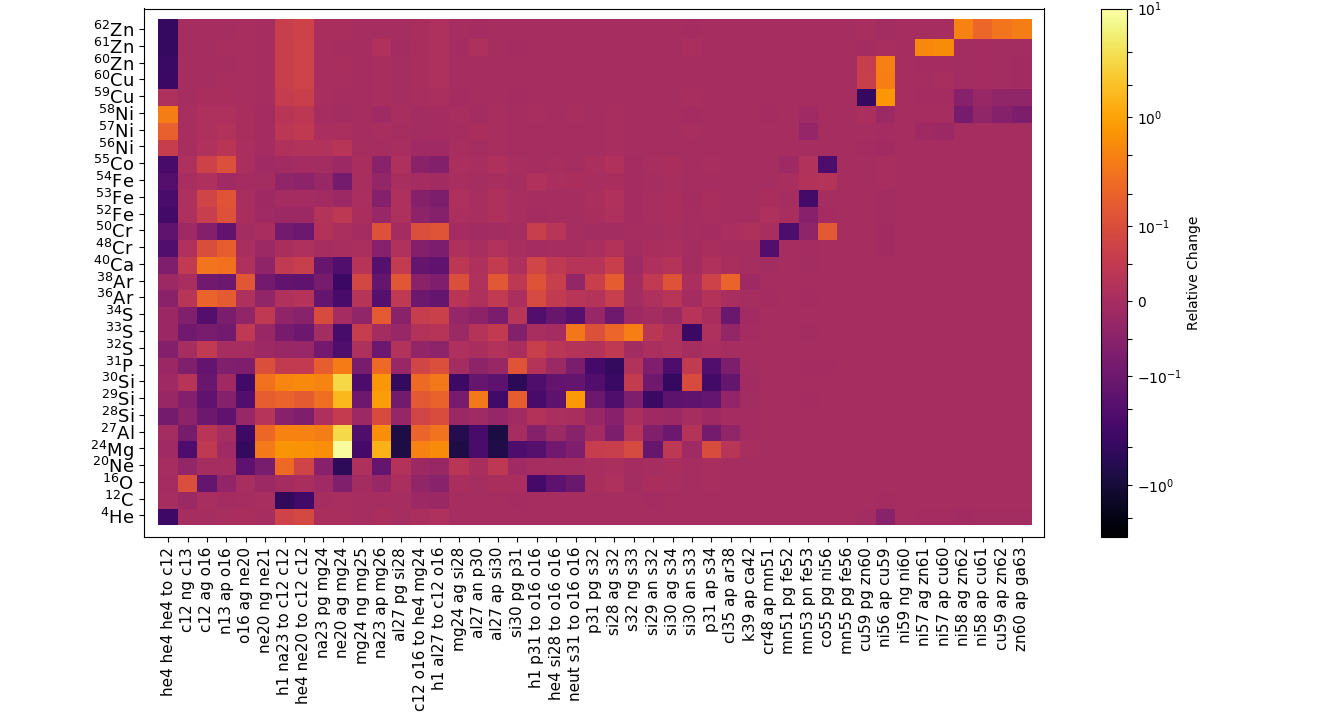}
  \label{10_up}
}

\subfloat[]{%
  \includegraphics[scale=.53]{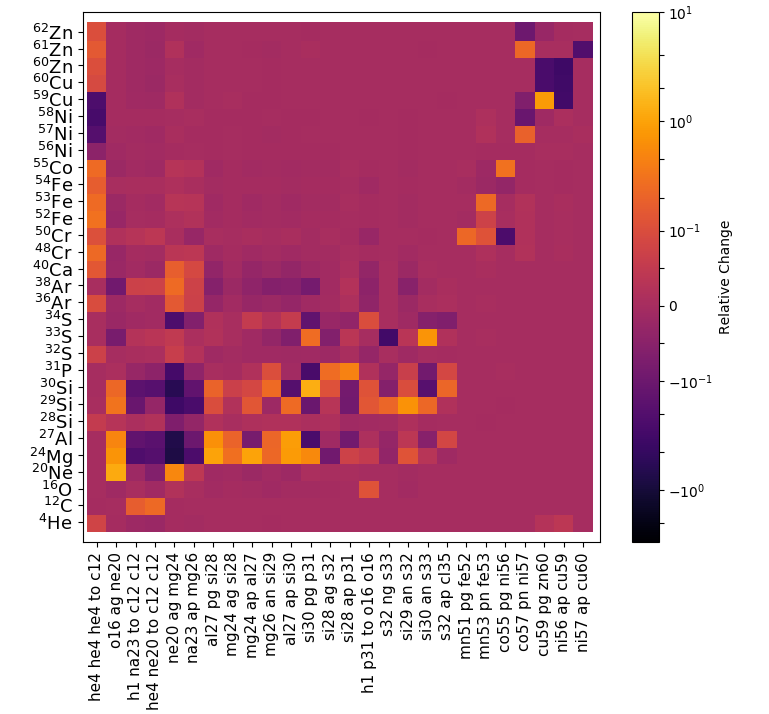}
  \label{10_dwn}
}

\caption{(a) The relative change in the nucleosynthetic yields generated with standard nuclear reaction rates normalized by a factor of 10.0 compared against the yields generated using standard nuclear reaction rates. The colorbar indicates the value of the relative change in logarithmic scale, and a small range of $\pm0.1$ around zero is mapped linearly to avoid ln(0). We present isotopes ($y$-axis) that achieve an initial abundance greater than $10^{-4}M_\odot$ and nuclear reactions ($x$-axis) which affected the yields of at least one species by $\geq 20\%$. We also include $^{27}$Al and $^{31}$P. (b) Same as Figure \ref{10_up}, but with standard reaction rates normalized by a factor of 0.1. The magnitude of the relative change is lower when nuclear reaction rates are normalized by a uniform factor of 0.1 as compared to using a factor of 10.0 normalization.}

\end{center}
\end{figure*}

\par
\begin{figure*}
\begin{center}
\subfloat[]{%
  \includegraphics[scale=.5]{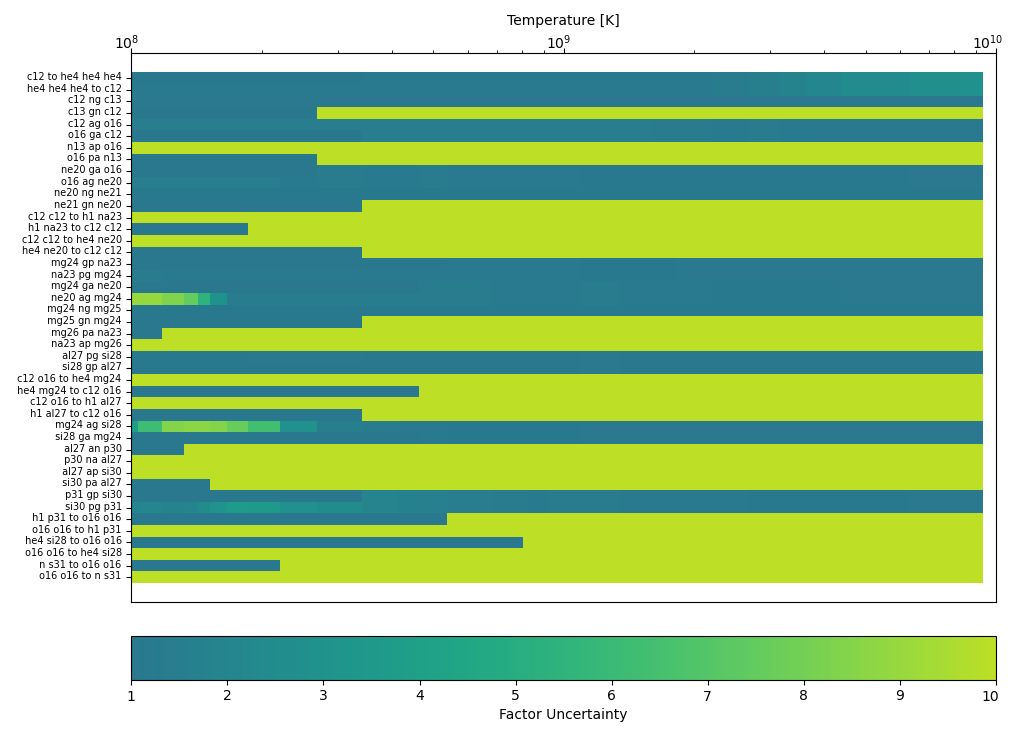}
  \label{uncer_1}
}

\subfloat[]{%
  \includegraphics[scale=.5]{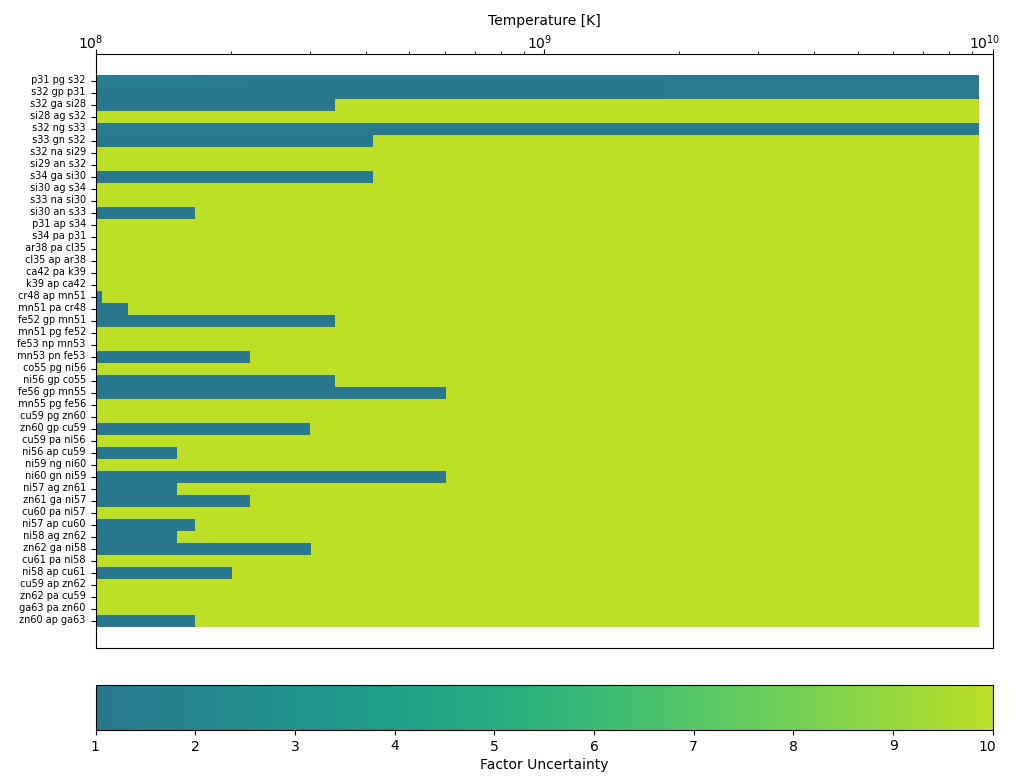}
  \label{uncer_2}
}

\caption{(a) The factor uncertainty of the rate as a function of temperature as provided by STARLIB for the first set of reactions, and their corresponding reverse processes, that induce a relative change of at least 20\% in the yield of at least one isotope in our nuclear network when normalized. (b) Same as Figure \ref{uncer_1}, but for the remaining reactions that induce a relative change of at least 20\% in the yield of at least one isotope in our nuclear network when normalized.}

\end{center}
\end{figure*}

 In Figure \ref{10_up} we show the relative change in isotopic abundances generated when each nuclear reaction rate in our network is individually normalized by a uniform factor of 10.0. Figure \ref{10_dwn} shows the results of an analogous calculation, but for the individual normalization of each nuclear reaction rate in our network by a uniform factor of 0.1. We apply these two normalizations to the rates of all nuclear reactions in our network, each time repeating all nucleosynthetic calculations, to identify those reactions whose uncertainty in the rate has the largest influence on the final chemical composition of the ejecta produced by the model. The normalization of all rates by a uniform factor of ten is an approximation to the actual temperature-dependent uncertainties of these reaction rates provided by STARLIB \citep{2013ApJS..207...18S} shown in Figures \ref{uncer_1} and \ref{uncer_2}. Most experimentally determined reaction rates follow a lognormal probability density, and the factor uncertainty of the rate is defined as $f.u. = e^{\sigma}$, where $\sigma$ is the lognormal spread parameter corresponding to the lognormal approximation of the rate. Only a select number of these reaction rates have been derived from Monte Carlo sampling of experimental data, and the remaining rates and uncertainties have been determined theoretically using Hauser-Feshbach model calculations \citep{2008PhRvC..78f4307G}. These theoretically determined rates are assigned a constant factor uncertainty $f.u. = 10$ at all temperatures. For a more detailed discussion of the calculation of nuclear reaction rate uncertainties see \citet{2013ApJS..207...18S}. For the purposes of our study, the normalization of all rates by a uniform factor of 10 is a sufficient approximation to explore the extent to which nucleosynthesis in the detonations of sub-$M_{\text{Ch}}$ WDs is sensitive to nuclear reaction rate uncertainties. For analogous studies employing a similar approach to investigate near-$M_{\text{Ch}}$ WD Type Ia explosion models, see \citet{2017ApJ...843...35M}, \citet{2013A&A...557A...3P}, and \citet{2012PhRvC..85e5805B}.

\begin{figure}[hbt!]
  \centering \includegraphics[scale=.57]{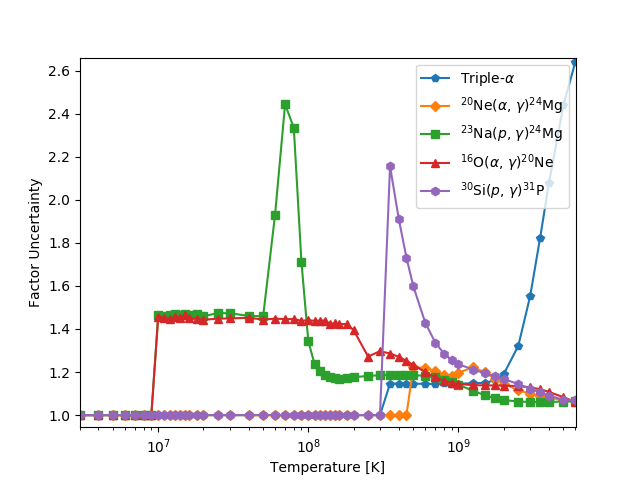}
  \caption{The temperature-dependent factor uncertainty of the rate provided by STARLIB for the triple-$\alpha$ process, $^{20}$Ne($\alpha$, $\gamma$)$^{24}$Mg, $^{23}$Na($p$, $\gamma$)$^{24}$Mg, $^{16}$O($\alpha$, $\gamma$)$^{20}$Ne, and $^{30}$Si($p$, $\gamma$)$^{31}$P, plotted over the temperatures reached during detonation.}
  \label{temp_uncer}
\end{figure}

\par From our nuclear reaction network of 942 nuclear reactions and 205 isotopes, we find that 50 reactions induce a change of at least $20\%$ on the final nucleosynthetic yield of at least one isotope. Uncertainties in the rates of $^{20}$Ne($\alpha$, $\gamma$)$^{24}$Mg, $^{23}$Na($p$, $\gamma$)$^{24}$Mg, $^{23}$Na($\alpha$, $p$)$^{26}$Mg, $^{16}$O($^{16}$O, $p$)$^{31}$P, $^{16}$O($\alpha$, $\gamma$)$^{20}$Ne, $^{30}$Si($p$, $\gamma$)$^{31}$P, and the triple-$\alpha$ process are the reactions which consistently have the most pronounced effects on nuclear yields. Figure \ref{temp_uncer} shows the temperature-dependent factor uncertainty for those reactions for which there is sufficient experimental data available for Monte Carlo based reaction rates and uncertainties, plotted over the temperatures reached during the detonation.

\subsection{Isotopic Ratios}
\par In this section, we analyze the impact of nuclear reaction rate variation on the model's predictions of important isotopic ratios. It is of interest to investigate the extent to which the results summarized in Figures \ref{10_up} and \ref{10_dwn} impact the model's predictions of $^{57}$Ni/$^{56}$Ni,  $^{55}$Co/$^{57}$Ni, and the final mass ratio of Mn/Fe after all radioactive decays have taken place. To account for decays, the masses of Mn and Fe are calculated as $^{55}$Mn + $^{55}$Co + $^{55}$Fe and $^{54}$Fe + $^{56}$Fe + $^{56}$Co + $^{56}$Ni + $^{57}$Fe + $^{57}$Co + $^{57}$Ni, respectively. The values of these ratios can serve as a useful tool for breaking the degeneracy between competing SN Ia progenitor models. Note that we calculate all isotopic ratios $10 \,$s after the start of explosion.

\begin{table}[]
\hspace{-14mm}
\scalebox{0.78}{
\begin{tabular}{lrccc}
\hline \hline
 & \multicolumn{1}{c|}{Reaction \& Normalization}                                                                                & \multicolumn{3}{c}{Isotopic Ratio}                                                                                                                                                                                    \\ \hline \hline 
 & \multicolumn{1}{r|}{}                                                                                                         & \multicolumn{1}{c|}{$^{57}$Ni/$^{56}$Ni}                                      & \multicolumn{1}{c|}{$^{55}$Co/$^{57}$Ni}                                  & Mn/Fe                                                     \\ \cline{1-5} 
 & \multicolumn{1}{r|}{\begin{tabular}[c]{@{}r@{}}triple-$\alpha$    $\times 10.0$\\ $\times 0.1$\end{tabular}}                  & \multicolumn{1}{c|}{\begin{tabular}[c]{@{}c@{}}12\%\\ -15\%\end{tabular}}     & \multicolumn{1}{c|}{\begin{tabular}[c]{@{}c@{}}-38\%\\ 53\%\end{tabular}} & \begin{tabular}[c]{@{}c@{}}-30\%\\ 31\%\end{tabular}      \\
 & \multicolumn{1}{r|}{\begin{tabular}[c]{@{}r@{}}$^{55}$Co($p$, $\gamma$)$^{56}$Ni    $\times10.0$\\ $\times 0.1$\end{tabular}} & \multicolumn{1}{c|}{\begin{tabular}[c]{@{}c@{}}$<$-1\%\\ $<$1\%\end{tabular}} & \multicolumn{1}{c|}{\begin{tabular}[c]{@{}c@{}}-23\%\\ 30\%\end{tabular}} & \begin{tabular}[c]{@{}c@{}}-23\%\\ 30\%\end{tabular}      \\
 & \multicolumn{1}{r|}{\begin{tabular}[c]{@{}r@{}}$^{57}$Co($p$, $n$)$^{57}$Ni    $\times 10.0$\\ $\times 0.1$\end{tabular}}     & \multicolumn{1}{c|}{\begin{tabular}[c]{@{}c@{}}-15\%\\ 19\%\end{tabular}}     & \multicolumn{1}{c|}{\begin{tabular}[c]{@{}c@{}}19\%\\ -16\%\end{tabular}} & \begin{tabular}[c]{@{}c@{}}$< $1\%\\ $<$-1\%\end{tabular}                                                                                                     \\ \hline \hline                                                      
\end{tabular}}
\caption{The relative change in the ratios of $^{57}$Ni/$^{56}$Ni, $^{55}$Co/$^{57}$Ni, and Mn/Fe generated by  applying two normalization factors (10.0$\times$ and 0.1$\times$) to the standard nuclear reaction rates.}
\label{ratio_impacts}
\end{table}

\par The effects that our nuclear reaction rate normalizations have on the values of these isotopic ratios are summarized in Table \ref{ratio_impacts}. In the standard case using default nuclear reaction rates, we obtain a value of 0.0155 for the ratio of $^{57}$Ni to $^{56}$Ni. In general, the value of this ratio was relatively robust to variations in nuclear reaction rates throughout our study. The ratio of $^{55}$Co to $^{57}$Ni achieves a value of 0.135 using standard nuclear reaction rates, and is affected by at most $53\%$ due to normalization of the triple-$\alpha$ process. The ratio of Mn/Fe achieves a value of $4.76 \times 10^{-3}$ using standard reaction rates, and changes by at most $31\%$ due to the normalization of the rate of the triple-$\alpha$ process.

\par Of these ratios, the value of $^{57}$Ni to $^{56}$Ni is the least sensitive to reaction rate normalizations. The final nucleosynthetic yields of these two isotopes are particularly robust to normalizations of nuclear reaction rates by a factor of 10.

\section{Conclusions}

\par Following up the work of \citet{2018ApJ...854...52S} who suggested the detonations of sub-$M_{\text{Ch}}$ WDs as promising candidates to explain most SNe Ia, but found enduring tensions in the neutron-rich nucleosynthesis generated in simulations of these explosions compared to SNe Ia observations, we have investigated the extent to which nuclear reaction rate uncertainties impact the nucleosynthetic yields of a one dimensional simulation of the detonation of a bare $1.0 \, M_{\odot}$ C/O WD.

\par To calculate nucleosynthetic yields, we post-process the results of the 41-isotope hydrodynamical simulation described in \citet{2018ApJ...854...52S} using a 205-isotope nuclear reaction network including all 942 interlinking reactions. To examine the impacts of nuclear reaction rate uncertainties we individually normalize the rates of all input nuclear reaction rates by two constant factors (10.0 and 0.1 times the standard rate), repeating all post-processing calculations for each individual normalization to a reaction rate and its corresponding reverse process.

\par Of the 942 nuclear reactions in our network, the individual normalization of 50 nuclear reaction rates impacts the nucleosynthetic yield of at least one isotope in our network by at least $20\%$ from the fiducial model. The rates of $^{20}$Ne($\alpha$, $\gamma$)$^{24}$Mg, $^{23}$Na($p$, $\gamma$)$^{24}$Mg, $^{23}$Na($\alpha$, $p$)$^{26}$Mg, $^{16}$O($^{16}$O, $p$)$^{31}$P, $^{16}$O($\alpha$, $\gamma$)$^{20}$Ne, $^{30}$Si($p$, $\gamma$)$^{31}$P, and the triple-$\alpha$ process have the greatest impact on supernova nucleosynthesis over all isotopes included in our nuclear network when normalized by a factor of 10.0 and 0.1. Individually increasing reaction rates by a factor of 10.0 has a more pronounced impact on nucleosynthetic yields than reducing these rates by the same factor. Individual normalizations to the rates of these reactions has the largest impact on the final abundances of isotopes in the range of 10 $\lesssim Z  \lesssim$ 20. In general, we find that the abundances of the Fe-group elements are relatively robust to individual normalizations of nuclear reaction rates in our network by constant factors of 10.

\par Our results are in general agreement with the previous studies of  \citet{2012PhRvC..85e5805B} and \citet{2013A&A...557A...3P}, who investigated the impacts of nuclear reaction rate uncertainties in simulations of near-$M_{\text{Ch}}$ WD SNe Ia scenarios including the W7 pure deflagration and the delayed detonation models. Both of these studies also found that the Fe-group elements were particularly robust to individual variations in the rates of nuclear reactions. This result can be explained by the fact that during explosion many of the Fe-group elements are produced in nuclear statistical equilibrium (NSE), which is insensitive to reaction rate variations. The nucleosynthetic yields in the explosions of WDs less massive than $1.0 \, M_{\odot}$ may be more sensitive to reaction rate uncertainties because more of the burning takes place outside of NSE. Future work is required to quantify the impact of nuclear reaction rate uncertainties on nucleosynthesis in lower-mass WD detonations.

\par We conclude that uncertainties in the rates of input nuclear reactions in the simulation of a bare $1.0 \, M_{{\odot}}$ C/O WD detonation have a small influence on the final chemical composition of the ejecta. Our results suggest that individual nuclear reaction rate uncertainties alone cannot account for current tensions between the isotopic ratios of Fe-group elements produced in simulations of sub-$M_{\text{Ch}}$ WD detonations and their values inferred through observations of late-time SNe Ia light curves, the Solar abundance of Mn, and SN remnant observations. Future work will be required to alleviate these tensions.

\section{Acknowledgements}
We thank Patrick J. Fitzpatrick, Alison Miller, and Dan Kasen for helpful discussions and advice. K.J.S. received support for this work from NASA through the Astrophysics Theory Program (NNX17AG28G). Calculations in this work were made using the Savio cluster provided by UCB.

\software{MESA version 8845.; \citet{2011ApJS..192....3P, 2013ApJS..208....4P, 2015ApJS..220...15P, 2018ApJS..234...34P}, 
FLASH version 4.2.2;  \citet{2000ApJS..131..273F, 2009arXiv0903.4875D}
 }

\clearpage
\bibliography{references}

\begin{thebibliography}{}
\expandafter\ifx\csname natexlab\endcsname\relax\def\natexlab#1{#1}\fi
\providecommand{\url}[1]{\href{#1}{#1}}
\providecommand{\dodoi}[1]{doi:~\href{http://doi.org/#1}{\nolinkurl{#1}}}
\providecommand{\doeprint}[1]{\href{http://ascl.net/#1}{\nolinkurl{http://ascl.net/#1}}}
\providecommand{\doarXiv}[1]{\href{https://arxiv.org/abs/#1}{\nolinkurl{https://arxiv.org/abs/#1}}}

\bibitem[{{Bildsten} {et~al.}(2007){Bildsten}, {Shen}, {Weinberg}, \&
  {Nelemans}}]{2007ApJ...662L..95B}
{Bildsten}, L., {Shen}, K.~J., {Weinberg}, N.~N., \& {Nelemans}, G. 2007, \apj,
  662, L95, \dodoi{10.1086/519489}

\bibitem[{{Bravo}(2019)}]{2019A&A...624A.139B}
{Bravo}, E. 2019, \aap, 624, A139, \dodoi{10.1051/0004-6361/201935095}

\bibitem[{{Bravo} {et~al.}(2019){Bravo}, {Badenes}, \&
  {Mart{\'\i}nez-Rodr{\'\i}guez}}]{2019MNRAS.482.4346B}
{Bravo}, E., {Badenes}, C., \& {Mart{\'\i}nez-Rodr{\'\i}guez}, H. 2019, \mnras,
  482, 4346, \dodoi{10.1093/mnras/sty2951}

\bibitem[{{Bravo} \& {Mart{\'{\i}}nez-Pinedo}(2012)}]{2012PhRvC..85e5805B}
{Bravo}, E., \& {Mart{\'{\i}}nez-Pinedo}, G. 2012, \prc, 85, 055805,
  \dodoi{10.1103/PhysRevC.85.055805}

\bibitem[{{Colgate} \& {McKee}(1969)}]{1969ApJ...157..623C}
{Colgate}, S.~A., \& {McKee}, C. 1969, \apj, 157, 623, \dodoi{10.1086/150102}

\bibitem[{{Cyburt} {et~al.}(2010){Cyburt}, {Amthor}, {Ferguson}, {Meisel},
  {Smith}, {Warren}, {Heger}, {Hoffman}, {Rauscher}, {Sakharuk}, {Schatz},
  {Thielemann}, \& {Wiescher}}]{2010ApJS..189..240C}
{Cyburt}, R.~H., {Amthor}, A.~M., {Ferguson}, R., {et~al.} 2010, The
  Astrophysical Journal Supplement Series, 189, 240,
  \dodoi{10.1088/0067-0049/189/1/240}

\bibitem[{De {et~al.}(2019)De, Kasliwal, Polin, Nugent, Bildsten, Adams, Bellm,
  Blagorodnova, Burdge, Cannella, \& et~al.}]{De_2019}
De, K., Kasliwal, M.~M., Polin, A., {et~al.} 2019, The Astrophysical Journal,
  873, L18, \dodoi{10.3847/2041-8213/ab0aec}

\bibitem[{{Dubey} {et~al.}(2009){Dubey}, {Reid}, {Weide}, {Antypas},
  {Ganapathy}, {Riley}, {Sheeler}, \& {Siegal}}]{2009arXiv0903.4875D}
{Dubey}, A., {Reid}, L.~B., {Weide}, K., {et~al.} 2009, arXiv e-prints,
  arXiv:0903.4875.
\newblock \doarXiv{0903.4875}

\bibitem[{{Fink} {et~al.}(2007){Fink}, {Hillebrandt}, \&
  {R{\"o}pke}}]{2007A&A...476.1133F}
{Fink}, M., {Hillebrandt}, W., \& {R{\"o}pke}, F.~K. 2007, \aap, 476, 1133,
  \dodoi{10.1051/0004-6361:20078438}

\bibitem[{{Fink} {et~al.}(2010){Fink}, {R{\"o}pke}, {Hillebrandt},
  {Seitenzahl}, {Sim}, \& {Kromer}}]{2010A&A...514A..53F}
{Fink}, M., {R{\"o}pke}, F.~K., {Hillebrandt}, W., {et~al.} 2010, \aap, 514,
  A53, \dodoi{10.1051/0004-6361/200913892}

\bibitem[{{Fryxell} {et~al.}(2000){Fryxell}, {Olson}, {Ricker}, {Timmes},
  {Zingale}, {Lamb}, {MacNeice}, {Rosner}, {Truran}, \&
  {Tufo}}]{2000ApJS..131..273F}
{Fryxell}, B., {Olson}, K., {Ricker}, P., {et~al.} 2000, \apjs, 131, 273,
  \dodoi{10.1086/317361}

\bibitem[{{Goriely} {et~al.}(2008){Goriely}, {Hilaire}, \&
  {Koning}}]{2008PhRvC..78f4307G}
{Goriely}, S., {Hilaire}, S., \& {Koning}, A.~J. 2008, \prc, 78, 064307,
  \dodoi{10.1103/PhysRevC.78.064307}

\bibitem[{{Graur} {et~al.}(2016){Graur}, {Zurek}, {Shara}, {Riess},
  {Seitenzahl}, \& {Rest}}]{2016ApJ...819...31G}
{Graur}, O., {Zurek}, D., {Shara}, M.~M., {et~al.} 2016, \apj, 819, 31,
  \dodoi{10.3847/0004-637X/819/1/31}

\bibitem[{{Hoeflich} \& {Khokhlov}(1996)}]{1996ApJ...457..500H}
{Hoeflich}, P., \& {Khokhlov}, A. 1996, \apj, 457, 500, \dodoi{10.1086/176748}

\bibitem[{{Homeier} {et~al.}(1998){Homeier}, {Koester}, {Hagen}, {Jordan},
  {Heber}, {Engels}, {Reimers}, \& {Dreizler}}]{1998A&A...338..563H}
{Homeier}, D., {Koester}, D., {Hagen}, H.~J., {et~al.} 1998, \aap, 338, 563

\bibitem[{{Kerzendorf} {et~al.}(2014){Kerzendorf}, {Taubenberger},
  {Seitenzahl}, \& {Ruiter}}]{2014ApJ...796L..26K}
{Kerzendorf}, W.~E., {Taubenberger}, S., {Seitenzahl}, I.~R., \& {Ruiter},
  A.~J. 2014, \apjl, 796, L26, \dodoi{10.1088/2041-8205/796/2/L26}

\bibitem[{{Kromer} {et~al.}(2010){Kromer}, {Sim}, {Fink}, {R{\"o}pke},
  {Seitenzahl}, \& {Hillebrandt}}]{2010ApJ...719.1067K}
{Kromer}, M., {Sim}, S.~A., {Fink}, M., {et~al.} 2010, \apj, 719, 1067,
  \dodoi{10.1088/0004-637X/719/2/1067}

\bibitem[{{Kushnir} {et~al.}(2020){Kushnir}, {Wygoda}, \&
  {Sharon}}]{2020MNRAS.499.4725K}
{Kushnir}, D., {Wygoda}, N., \& {Sharon}, A. 2020, \mnras, 499, 4725,
  \dodoi{10.1093/mnras/staa3017}

\bibitem[{{Livne}(1990)}]{1990ApJ...354L..53L}
{Livne}, E. 1990, \apjl, 354, L53, \dodoi{10.1086/185721}

\bibitem[{{Maoz} {et~al.}(2014){Maoz}, {Mannucci}, \&
  {Nelemans}}]{2014ARA&A..52..107M}
{Maoz}, D., {Mannucci}, F., \& {Nelemans}, G. 2014, Annual Review of Astronomy
  and Astrophysics, 52, 107, \dodoi{10.1146/annurev-astro-082812-141031}

\bibitem[{{Mart{\'\i}nez-Rodr{\'\i}guez}
  {et~al.}(2017){Mart{\'\i}nez-Rodr{\'\i}guez}, {Badenes}, {Yamaguchi},
  {Bravo}, {Timmes}, {Miles}, {Townsley}, {Piro}, {Mori}, {Andrews}, \&
  {Park}}]{2017ApJ...843...35M}
{Mart{\'\i}nez-Rodr{\'\i}guez}, H., {Badenes}, C., {Yamaguchi}, H., {et~al.}
  2017, \apj, 843, 35, \dodoi{10.3847/1538-4357/aa72f8}

\bibitem[{{Miles} {et~al.}(2018){Miles}, {Townsley}, {Shen}, {Timmes}, \&
  {Moore}}]{2018arXiv180607820M}
{Miles}, B.~J., {Townsley}, D.~M., {Shen}, K.~J., {Timmes}, F.~X., \& {Moore},
  K. 2018, arXiv e-prints.
\newblock \doarXiv{1806.07820}

\bibitem[{{Nomoto}(1982{\natexlab{a}})}]{1982ApJ...257..780N}
{Nomoto}, K. 1982{\natexlab{a}}, \apj, 257, 780, \dodoi{10.1086/160031}

\bibitem[{{Nomoto}(1982{\natexlab{b}})}]{1982ApJ...253..798N}
---. 1982{\natexlab{b}}, \apj, 253, 798, \dodoi{10.1086/159682}

\bibitem[{{Nomoto} {et~al.}(1984){Nomoto}, {Thielemann}, \&
  {Yokoi}}]{1984ApJ...286..644N}
{Nomoto}, K., {Thielemann}, F.-K., \& {Yokoi}, K. 1984, \apj, 286, 644,
  \dodoi{10.1086/162639}

\bibitem[{{Nugent} {et~al.}(1997){Nugent}, {Baron}, {Branch}, {Fisher}, \&
  {Hauschildt}}]{1997ApJ...485..812N}
{Nugent}, P., {Baron}, E., {Branch}, D., {Fisher}, A., \& {Hauschildt}, P.~H.
  1997, \apj, 485, 812, \dodoi{10.1086/304459}

\bibitem[{{Pankey}(1962)}]{1962PhDT........25P}
{Pankey}, Jr., T. 1962, PhD thesis, HOWARD UNIVERSITY.

\bibitem[{{Parikh} {et~al.}(2013){Parikh}, {Jos{\'e}}, {Seitenzahl}, \&
  {R{\"o}pke}}]{2013A&A...557A...3P}
{Parikh}, A., {Jos{\'e}}, J., {Seitenzahl}, I.~R., \& {R{\"o}pke}, F.~K. 2013,
  \aap, 557, A3, \dodoi{10.1051/0004-6361/201321518}

\bibitem[{{Paxton} {et~al.}(2011){Paxton}, {Bildsten}, {Dotter}, {Herwig},
  {Lesaffre}, \& {Timmes}}]{2011ApJS..192....3P}
{Paxton}, B., {Bildsten}, L., {Dotter}, A., {et~al.} 2011, \apjs, 192, 3,
  \dodoi{10.1088/0067-0049/192/1/3}

\bibitem[{{Paxton} {et~al.}(2013){Paxton}, {Cantiello}, {Arras}, {Bildsten},
  {Brown}, {Dotter}, {Mankovich}, {Montgomery}, {Stello}, {Timmes}, \&
  {Townsend}}]{2013ApJS..208....4P}
{Paxton}, B., {Cantiello}, M., {Arras}, P., {et~al.} 2013, \apjs, 208, 4,
  \dodoi{10.1088/0067-0049/208/1/4}

\bibitem[{{Paxton} {et~al.}(2015){Paxton}, {Marchant}, {Schwab}, {Bauer},
  {Bildsten}, {Cantiello}, {Dessart}, {Farmer}, {Hu}, {Langer}, {Townsend},
  {Townsley}, \& {Timmes}}]{2015ApJS..220...15P}
{Paxton}, B., {Marchant}, P., {Schwab}, J., {et~al.} 2015, \apjs, 220, 15,
  \dodoi{10.1088/0067-0049/220/1/15}

\bibitem[{{Paxton} {et~al.}(2018){Paxton}, {Schwab}, {Bauer}, {Bildsten},
  {Blinnikov}, {Duffell}, {Farmer}, {Goldberg}, {Marchant}, {Sorokina},
  {Thoul}, {Townsend}, \& {Timmes}}]{2018ApJS..234...34P}
{Paxton}, B., {Schwab}, J., {Bauer}, E.~B., {et~al.} 2018, \apjs, 234, 34,
  \dodoi{10.3847/1538-4365/aaa5a8}

\bibitem[{{Perlmutter} {et~al.}(1999){Perlmutter}, {Aldering}, {Goldhaber},
  {Knop}, {Nugent}, {Castro}, {Deustua}, {Fabbro}, {Goobar}, {Groom}, {Hook},
  {Kim}, {Kim}, {Lee}, {Nunes}, {Pain}, {Pennypacker}, {Quimby}, {Lidman},
  {Ellis}, {Irwin}, {McMahon}, {Ruiz-Lapuente}, {Walton}, {Schaefer}, {Boyle},
  {Filippenko}, {Matheson}, {Fruchter}, {Panagia}, {Newberg}, {Couch}, \&
  {Project}}]{1999ApJ...517..565P}
{Perlmutter}, S., {Aldering}, G., {Goldhaber}, G., {et~al.} 1999, \apj, 517,
  565, \dodoi{10.1086/307221}

\bibitem[{{Riess} {et~al.}(1998){Riess}, {Filippenko}, {Challis},
  {Clocchiatti}, {Diercks}, {Garnavich}, {Gilliland}, {Hogan}, {Jha},
  {Kirshner}, {Leibundgut}, {Phillips}, {Reiss}, {Schmidt}, {Schommer},
  {Smith}, {Spyromilio}, {Stubbs}, {Suntzeff}, \&
  {Tonry}}]{1998AJ....116.1009R}
{Riess}, A.~G., {Filippenko}, A.~V., {Challis}, P., {et~al.} 1998, \aj, 116,
  1009, \dodoi{10.1086/300499}

\bibitem[{{R{\"o}pke} {et~al.}(2012){R{\"o}pke}, {Kromer}, {Seitenzahl},
  {Pakmor}, {Sim}, {Taubenberger}, {Ciaraldi-Schoolmann}, {Hillebrandt},
  {Aldering}, {Antilogus}, {Baltay}, {Benitez-Herrera}, {Bongard}, {Buton},
  {Canto}, {Cellier-Holzem}, {Childress}, {Chotard}, {Copin}, {Fakhouri},
  {Fink}, {Fouchez}, {Gangler}, {Guy}, {Hachinger}, {Hsiao}, {Chen},
  {Kerschhaggl}, {Kowalski}, {Nugent}, {Paech}, {Pain}, {Pecontal}, {Pereira},
  {Perlmutter}, {Rabinowitz}, {Rigault}, {Runge}, {Saunders}, {Smadja},
  {Suzuki}, {Tao}, {Thomas}, {Tilquin}, \& {Wu}}]{2012ApJ...750L..19R}
{R{\"o}pke}, F.~K., {Kromer}, M., {Seitenzahl}, I.~R., {et~al.} 2012, \apj,
  750, L19, \dodoi{10.1088/2041-8205/750/1/L19}

\bibitem[{{Sallaska} {et~al.}(2013){Sallaska}, {Iliadis}, {Champange},
  {Goriely}, {Starrfield}, \& {Timmes}}]{2013ApJS..207...18S}
{Sallaska}, A.~L., {Iliadis}, C., {Champange}, A.~E., {et~al.} 2013, The
  Astrophysical Journal Supplement Series, 207, 18,
  \dodoi{10.1088/0067-0049/207/1/18}

\bibitem[{{Seitenzahl} {et~al.}(2013){Seitenzahl}, {Cescutti}, {R{\"o}pke},
  {Ruiter}, \& {Pakmor}}]{2013A&A...559L...5S}
{Seitenzahl}, I.~R., {Cescutti}, G., {R{\"o}pke}, F.~K., {Ruiter}, A.~J., \&
  {Pakmor}, R. 2013, \aap, 559, L5, \dodoi{10.1051/0004-6361/201322599}

\bibitem[{{Shappee} {et~al.}(2017){Shappee}, {Stanek}, {Kochanek}, \&
  {Garnavich}}]{2017ApJ...841...48S}
{Shappee}, B.~J., {Stanek}, K.~Z., {Kochanek}, C.~S., \& {Garnavich}, P.~M.
  2017, \apj, 841, 48, \dodoi{10.3847/1538-4357/aa6eab}

\bibitem[{{Shen} \& {Bildsten}(2014)}]{2014ApJ...785...61S}
{Shen}, K.~J., \& {Bildsten}, L. 2014, \apj, 785, 61,
  \dodoi{10.1088/0004-637X/785/1/61}

\bibitem[{{Shen} {et~al.}(2021{\natexlab{a}}){Shen}, {Blondin}, {Kasen},
  {Dessart}, {Townsley}, {Boos}, \& {Hillier}}]{2021ApJ...909L..18S}
{Shen}, K.~J., {Blondin}, S., {Kasen}, D., {et~al.} 2021{\natexlab{a}}, \apjl,
  909, L18, \dodoi{10.3847/2041-8213/abe69b}

\bibitem[{{Shen} {et~al.}(2021{\natexlab{b}}){Shen}, {Boos}, {Townsley}, \&
  {Kasen}}]{2021arXiv210812435S}
{Shen}, K.~J., {Boos}, S.~J., {Townsley}, D.~M., \& {Kasen}, D.
  2021{\natexlab{b}}, arXiv e-prints, arXiv:2108.12435.
\newblock \doarXiv{2108.12435}

\bibitem[{{Shen} {et~al.}(2018{\natexlab{a}}){Shen}, {Kasen}, {Miles}, \&
  {Townsley}}]{2018ApJ...854...52S}
{Shen}, K.~J., {Kasen}, D., {Miles}, B.~J., \& {Townsley}, D.~M.
  2018{\natexlab{a}}, \apj, 854, 52, \dodoi{10.3847/1538-4357/aaa8de}

\bibitem[{{Shen} \& {Moore}(2014)}]{2014ApJ...797...46S}
{Shen}, K.~J., \& {Moore}, K. 2014, \apj, 797, 46,
  \dodoi{10.1088/0004-637X/797/1/46}

\bibitem[{{Shen} {et~al.}(2018{\natexlab{b}}){Shen}, {Boubert}, {G{\"a}nsicke},
  {Jha}, {Andrews}, {Chomiuk}, {Foley}, {Fraser}, {Gromadzki}, {Guillochon},
  {Kotze}, {Maguire}, {Siebert}, {Smith}, {Strader}, {Badenes}, {Kerzendorf},
  {Koester}, {Kromer}, {Miles}, {Pakmor}, {Schwab}, {Toloza}, {Toonen},
  {Townsley}, \& {Williams}}]{2018ApJ...865...15S}
{Shen}, K.~J., {Boubert}, D., {G{\"a}nsicke}, B.~T., {et~al.}
  2018{\natexlab{b}}, \apj, 865, 15, \dodoi{10.3847/1538-4357/aad55b}

\bibitem[{{Thielemann} {et~al.}(1986){Thielemann}, {Nomoto}, \&
  {Yokoi}}]{1986A&A...158...17T}
{Thielemann}, F.-K., {Nomoto}, K., \& {Yokoi}, K. 1986, \aap, 158, 17

\bibitem[{{Timmes} {et~al.}(1995){Timmes}, {Woosley}, \&
  {Weaver}}]{1995ApJS...98..617T}
{Timmes}, F.~X., {Woosley}, S.~E., \& {Weaver}, T.~A. 1995, \apjs, 98, 617,
  \dodoi{10.1086/192172}

\bibitem[{Truran {et~al.}(1967)Truran, Arnett, \&
  Cameron}]{doi:10.1139/p67-184}
Truran, J.~W., Arnett, W.~D., \& Cameron, A. G.~W. 1967, Canadian Journal of
  Physics, 45, 2315, \dodoi{10.1139/p67-184}

\bibitem[{{Webbink}(1984)}]{1984ApJ...277..355W}
{Webbink}, R.~F. 1984, \apj, 277, 355, \dodoi{10.1086/161701}

\bibitem[{{Whelan} \& {Iben}(1973)}]{1973ApJ...186.1007W}
{Whelan}, J., \& {Iben}, Jr., I. 1973, \apj, 186, 1007, \dodoi{10.1086/152565}

\bibitem[{{Woosley} \& {Kasen}(2011)}]{2011ApJ...734...38W}
{Woosley}, S.~E., \& {Kasen}, D. 2011, \apj, 734, 38,
  \dodoi{10.1088/0004-637X/734/1/38}

\bibitem[{{Woosley} \& {Weaver}(1994)}]{1994ApJ...423..371W}
{Woosley}, S.~E., \& {Weaver}, T.~A. 1994, \apj, 423, 371,
  \dodoi{10.1086/173813}

\end{thebibliography}
\end{document}